\documentclass[12pt]{article}
\usepackage{graphicx}
\usepackage{longtable}

\begin{document}

\makeatletter
\renewcommand*{\@cite}[2]{{#2}}
\renewcommand*{\@biblabel}[1]{#1.\hfill}
\makeatother

\title{Pulkovo Compilation of Radial Velocities for 35495 Hipparcos Stars in a Common System}
\author{G.~A.~Gontcharov\thanks{E-mail: georgegontcharov@yahoo.com}}

\maketitle

Pulkovo Astronomical Observatory, Russian Academy of Sciences, Pulkovskoe sh. 65, St. Petersburg, 196140 Russia


The Pulkovo Compilation of Radial Velocities (PCRV) has been made to study the stellar
kinematics in the local spiral arm. The PCRV contains weighted mean absolute radial velocities for 35 495
Hipparcos stars of various spectral types and luminosity classes over the entire celestial sphere mainly
within 500 pc of the Sun. The median accuracy of the radial velocities obtained is 0.7 km s$^{-1}$. Results from
203 publications were used in the catalogue. Four of them were used to improve the radial velocities of
standard stars from the IAU list. The radial velocities of 155 standard stars turned out to be constant within
0.3 km s$^{-1}$. These stars were used to analyze 47 768 mean radial velocities for 37 200 stars from 12 major
publications ($\sim80\%$ of all the data used). Zero-point discrepancies and systematic dependences on radial
velocity, $(B-V)$ color index, right ascension, and declination were found in radial velocity differences of the
form ``publication minus IAU list of standards''. These discrepancies and dependences were approximated
and taken into account when calculating the weighted mean radial velocities. 1128 stars whose independent
radial-velocity determinations were available at least in three of these publications and agreed within
3 km s$^{-1}$ were chosen as the work list of secondary standards. Radial-velocity differences of the form
``publication minus list of secondary standards'' were used by analogy to correct the zero points and
systematic dependences in the radial velocities from 33 more publications (~13\% of the data used). In
addition, the radial velocities from 154 minor publications ($\sim7\%$ of the data used) pertaining to well-known
instruments were used without any corrections.

\newpage
\section*{INTRODUCTION}

The stellar kinematics in the solar neighborhood
within the local spiral arm of the Galaxy is being
studied at the Pulkovo Astronomical Observatory
of the Russian Academy of Sciences and the Astronomical
Institute of the St. Petersburg State
University as part of the OSACA (Orion Spiral Arm
Catalogue, http://www.geocities.com/orionspiral/)
project (Gontcharov 2004). The study is based on
six quantities that describe the stellar positions and
velocities: $\alpha$, $\delta$, $\pi$, $\mu_{\alpha}$, $\mu_{\delta}$, and $V_r$, as well as the
coordinates $X$, $Y$ , and $Z$ and space velocities $U$, $V$ ,
and $W$ calculated from them in the standard Galactic
rectangular coordinate system, where $X$, $Y$, and
$Z$ increase in the directions of the Galactic center,
the Galactic rotation, and the Galactic North Pole,
respectively.

At present, the coordinates and proper motions are
known for a larger number of stars and with a higher
relative accuracy than the parallaxes and radial velocities.
For example, the coordinates and proper motions
of $\sim96 000$ Hipparcos stars (ESA 1997) within
500 pc of the Sun are known with a relative accuracy
higher than $10^{-7}$ and an accuracy higher than
3 km s$^{-1}$, respectively, while the parallaxes of only
$\sim69 000$ Hipparcos stars are known with a relative
accuracy higher than 0.3 and the radial velocities
of only $\sim20 000$ stars from the WEB (Duflot et al.
1995b) and Barbier-Brossat and Figon (below referred
to as BBF) (2000) catalogues are known with
an accuracy higher than 5 km s$^{-1}$.

Several major \emph{observational} catalogues of radial
velocities, including the Geneva-Copenhagen survey
of radial velocities for more than 14 000 stars in
the solar neighborhood (below referred to as GCS)
(Nordstr\"om et al. 2004) and the kinematic survey of
more than 6000 K and M giants based on observations
with CORAVEL spectrometers (below referred
to as KMG) (Famaey et al. 2005), have been
compiled since the publication of these compilation
catalogs. Thus, making a new compilation of radial
velocities presented in this paper has become a necessity.
By combining data from the WEB, BBF, GCS,
and KMG catalogues and 199 more catalogues of
radial velocities, we have made the Pulkovo Compilation
of Radial Velocities (PCRV) for 35 495 Hipparcos
stars to study their kinematics. Supplementing
the PCRV by new observations and improving the
weighted mean radial velocities are an important part
of the long-term OSACA project. The size of this
paper is too small to describe the kinematic studies
using the PCRV that have been considered in other
publications, for example, by Bobylev et al. (2006).

\section*{COMBINING THE RADIAL VELOCITIES FROM VARIOUS PUBLICATIONS}

The radial-velocity measurements performed with
different instruments have different systematic errors.
In contrast to the relative radial-velocity measurements
to detect extrasolar planets, the difficulties in
determining the absolute radial velocities for stellar
kinematics, i.e., the velocities relative to the barycenter
of the Solar System, are particularly great. The
systematic errors in measuring the absolute radial
velocities and the attainable level of accuracy were
discussed, for example, by Nidever et al. (2002). They
showed that apart from well-known effects, for example,
the Earth's motion around the Sun, of greatest
importance for the absolute radial velocities are the
gravitational redshift of spectral lines in the stellar
gravitational field, the convective blueshift of lines in
the upper stellar layer, and the difference between
the observed and comparison spectra. According to
the calculations by Nidever et al. (2002), partly compensating
each other, these effects do not exceed
1 km s$^{-1}$ and must be approximated by linear or
quadratic dependences on the color index or the radial
velocity itself. Such dependences are commonly
found in the radial-velocity differences from various
publications.

The new astrometric methods of radial-velocity
measurements that are free from the above-mentioned
systematic errors (Madsen et al. 2002) are promising.
However, they can be implemented with acceptable
accuracy and massiveness only in future
space astrometry missions. The catalogue by Madsen
et al. (2002) was not used to make the PCRV, since
the radial-velocity differences of the form ``astrometric
minus spectroscopic'' were found to depend significantly
on parallax, which requires a detailed analysis.

Traditionally, when the radial velocities are measured,
the positions of spectral lines in the observed
spectrum and the spectrum of a standard star with a
constant and exactly known radial velocity are compared.
The International Astronomical Union (IAU)
has approved a list of standard stars. Over several
decades, this list, on the one hand, has been supplemented
and, on the other hand, stars with variable
radial velocities have been excluded from it on
the basis of regularly repeated measurements. The
current observations of IAU standard stars were presented,
for example, by Stefanik et al. (1999), Udry
et al. (1999a, 1999b), and Fekel (1999). After these
data have been averaged and the known binary and
variable stars have been excluded, I have used the
remaining 155 stars with radial velocities constant
within 0.3 km s$^{-1}$ for many years as primary standards
(below referred to as the IAU list of standards).
These stars are presented in Table 1: it gives their
Hipparcos and HD numbers, mean radial velocities
with their accuracies $\sigma(V_r)$ from the four mentioned
publications (in km s$^{-1}$), approximate $\alpha$ and $\delta$ in
degrees with fractions (J2000), and $V$ magnitudes.
These stars do not represent the whole variety of
stellar classes and cover nonuniformly the celestial
sphere. Therefore, standard stars are difficult to select
for many observational programs and the compiled
catalogs are difficult to compare with other catalogs.
Thus, expanding the list of standard stars or
compiling a list of secondary standards, especially by
incorporating faint stars, with the inclusion of stars
of all spectral types and luminosity classes uniformly
distributed over the celestial sphere, are of current
interest.

\begin{table}
\caption[]{IAU standard stars used to make the PCRV (the
full table is available in electronic form).}
\label{iau}
\[
\begin{tabular}{rrrrrrr}
\hline
\noalign{\smallskip}
 HIP & HD & $V_r$ & $\sigma(V_r)$ & $\alpha$ & $\delta$ & $m_V$ \\
\hline
\noalign{\smallskip}
   910 &   693 &  14.4 & 0.1 &  2.816 & -15.468 & 4.89 \\
  1499 &  1461 & -10.2 & 0.1 &  4.674 &  -8.053 & 6.47 \\
  1803 &  1835 &  -2.6 & 0.1 &  5.716 & -12.209 & 6.39 \\
  2920 &  3360 &  -0.3 & 0.1 &  9.243 &  53.897 & 3.69 \\
  3179 &  3712 &  -4.3 & 0.1 & 10.127 &  56.537 & 2.24 \\
\hline
\end{tabular}
\]
\end{table}

Twelve major publications contain the observations
of the above-mentioned IAU standard stars that
are enough to estimate and correct the zero points
and systematic errors of the radial velocities in these
publications. For these publications, Table 2 gives a
bibliographic code of the publication in the SIMBAD
database, (http://simbad.u-strasbg.fr/sim-fid.pl) a
reference to the publication, the number of stars
from the publication used in the PCRV (N), and
the number of IAU standards used (n). A total of
47 768 mean radial velocities for 37 200 stars were
taken from these publications (but, in the end, some
of the spectroscopic binaries were not included in the
PCRV). These account for $\sim80\%$ of the catalogued
mean radial velocities used to make the PCRV.

For the IAU standards, we calculated radialvelocity
differences of the form ``publication minus
IAU list of standards''. The mean of these differences
presented in the $\overline{\Delta V_r}$ column (in km s$^{-1}$) of Table 2
is considered as a zero-point discrepancy between
the publication and the IAU list of standards. The
standard deviation of the differences presented in the
$\sigma$ column (in km s$^{-1}$) is considered as an objective estimate
of the mean radial-velocity accuracy in a given
publication. The significant and physically justified
systematic dependences of the differences on various
factors are considered as systematic radial-velocity
errors in a given publication. They are approximated
by the equations presented in Table 3 with the opposite
sign (in km s$^{-1}$). In the correction formulas,
$\alpha$ and $\delta$ are in degrees, the radial velocity $V_r$ is in
km s$^{-1}$, and the color index $(B-V)$ is in magnitudes.
After the corrections to the radial velocities have been
applied, residual differences of the form ``publication
minus IAU list of standards'' are calculated. The
standard deviation of these differences presented in
the $\sigma'$ column of Table 2 is considered as an objective
estimate of the mean radial-velocity accuracy after
the correction. The latter is compared with the mean
accuracy declared by the authors of the publication
for stars of the same magnitude. For all the 12 publications,
these accuracies are in good agreement.
Therefore, the accuracy declared by the authors of the
publications (changing within the publication from
star to star) was used to calculate the weighted mean
radial velocities and their accuracies: the weights
were assigned as $1/(\sigma_0^2+\sigma_1^2+0.2^2)$, where $\sigma_0$ is the
accuracy declared by the authors of the publication
(in km s$^{-1}$) and $\sigma_1$ is an estimate of the error due to
uncertainty in the approximation factors (in km s$^{-1}$).
The value of 0.2 km s$^{-1}$ was added to eliminate the
unjustifiably great weights in the cases where the
formally declared accuracy is less than 0.3 km s$^{-1}$.
Among the approximation factors, $\alpha$ and $\delta$ were taken
from the Hipparcos catalogue and may be considered
accurate, $(B-V)$ from Hipparcos for the stars under
consideration is known with an accuracy higher than $0.1^m$,
and $V_r$ is known for the standard stars with an
accuracy higher than 0.3 km s$^{-1}$.As a result, $\sigma_1$ does
not exceed 1 km s$^{-1}$.

\begin{table}
\caption[]{Radial-velocity publications with a sufficient number of IAU standard stars.}
\label{pubiau}
\[
\begin{tabular}{llrrrrr}
\hline
\noalign{\smallskip}
  CDS bibcode   &   Reference & N   &  n & $\overline{\Delta V_r}$ & $\sigma$ & $\sigma'$ \\
\hline
\noalign{\smallskip}
2004A\&A...418..989N & Nordstr\"om et al. (2004)        & 11545 & 59 & $-0.2$ & 0.3 & 0.1 \\
1995A\&AS..114..269D & Duflot et al. (1995b)            & 11291 & 58 & $+0.3$ & 0.9 & 0.9 \\
2000A\&AS..142..217B & Barbier-Brossat and Figon (2000) & 11193 & 86 & $+0.0$ & 1.7 & 1.7 \\
2005A\&A...430..165F & Famaey et al. (2005)             &  5406 &  8 & $+0.0$ & 0.2 & 0.2 \\
1999A\&AS..137..451G & Grenier et al. (1999b)           &  2876 &  9 & $+0.7$ & 4.2 & 2.0 \\
1999A\&AS..139..433D & deMedeiros and Mayor (1999)      &  1414 & 29 & $-0.3$ & 0.2 & 0.1 \\
2004PASP..116..693M & Moultaka et al. (2004)           &  1248 & 65 & $+0.0$ & 0.5 & 0.4 \\
2000A\&AS..142..275S & Strassmeier et al. (2000)        &  1087 & 11 & $+0.3$ & 0.7 & 0.6 \\
2002ApJS..141..503N & Nidever et al. (2002)            &   849 & 54 & $+0.1$ & 0.1 & 0.1 \\ 
1994AJ....107.2240C & Carney et al. (1994)             &   773 &  8 & $-0.2$ & 0.5 & 0.2 \\
1991AJ....101.1495M & Morse et al. (1991)              &    61 & 10 & $-0.6$ & 1.2 & 1.0 \\
2000PASP..112..966S & Skuljan et al. (2000)            &    25 & 14 & $+0.1$ & 0.2 & 0.1 \\
\hline
\end{tabular}
\]
\end{table}

\begin{table}
\def\baselinestretch{1}\normalsize\scriptsize
\caption[]{Systematic radial-velocity corrections for the publications with a sufficient number of IAU standard stars.}
\label{coriau}
\[
\begin{tabular}{lc}
\hline
\noalign{\smallskip}
  References   & Correction \\
\hline
\noalign{\smallskip}
Nordstr\"om et al. (2004)        & $-2.52(B-V)^2+4.7(B-V)-0.002V_r+0.0006\alpha+0.04\cos(\alpha-40)-1.88$ \\ 
Duflot et al. (1995b)            & $0$ \\
Barbier-Brossat and Figon (2000) & $0$ \\
Famaey et al. (2005)             & $0$ \\
Grenier et al. (1999b)           & $-0.231\delta+5.52$ \\
deMedeiros and Mayor (1999)      & $-1.23(B-V)^2+2.9(B-V)-0.002V_r+0.0004\alpha+0.06\cos(\alpha+20)-1.30$ \\ 
Moultaka et al. (2004)           & $+0.84(B-V)-0.6\qquad\mbox{ïðè }B-V<0.4, +0.87(B-V)-0.9\qquad\mbox{ïðè }B-V>1.1$ \\
Strassmeier et al. (2000)        & $-0.5\cos(2(\alpha+10))-0.2$ \\
Nidever et al. (2002)            & $+0.12(B-V)-0.001V_r-0.2$ \\ 
Carney et al. (1994)             & $-0.85(B-V)-0.0038V_r+0.58$ \\
Morse et al. (1991)              & $-0.036V_r-0.77\cos(\alpha)-0.023\delta+1.56$ \\
Skuljan et al. (2000)            & $+0.2016(B-V)+0.0029V_r-0.3$ \\
\hline
\end{tabular}
\]
\end{table}

Initially, we considered the dependences of the
radial-velocity differences on the radial velocity itself,
equatorial and Galactic coordinates, parallax,
$V$ magnitude, and $(B-V)$ color index taken from
Hipparcos, and absolute magnitude $M_V$ calculated
here from Hipparcos data by applying the correction
for interstellar extinction as prescribed by Arenou
et al. (1992). For checking purposes, we also considered
the dependences on stellar rotational velocity
$v \sin i$, metallicity $Fe/H$, age, effective temperature,
$b-y$ color index in Str\"omgren photometry, and other
quantities taken from the publications under consideration
or from the GCS, where many of these
quantities were determined with a high accuracy for
thousands of stars, but not for all stars. Therefore,
although slight dependences on $v \sin i$, $Fe/H$, and age
were found in several cases, they were disregarded
and will be analyzed at a later time.

A multifactor regression analysis was used to analyze
the dependences. Significant dependences were
sought at a multiple correlation coefficient larger than
0.6 and pair correlation coefficients larger than 0.4.
The factors were considered in order of decreasing
pair correlation coefficient. If the number of stars is
much larger than the number of factors, then a dependence
that is formally significant according to the
F test will always be found. Therefore, it is particularly
important to physically justify the dependences
found and to exclude the factors that correlate with
one another from the analysis.

As expected, a strong correlation was found between
the effective temperature and the $(B-V)$ and
$(b-y)$ color indices. Moreover, for the publications that
include only main-sequence stars or only giant stars,
these quantities correlate with the absolute magnitude $M_V$,
since the main sequence itself and the giant
branch reflect the color-magnitude relation for stars.
In many cases, the dependence on $(b-y)$ was more
pronounced than that on $(B-V)$. However, $(B-V)$ was
chosen by tradition and, besides, it is known for all of
the stars under consideration.

A dependence of the radial velocity on hour angle
is possible in Coude observations. In some of
the observational programs, the hour angle correlates
with $\alpha$, while the relationship between hour angle
and atmospheric dispersion depends on $\delta$. The linear
and sinusoidal dependences of the radial velocity on
equatorial coordinates can be explained in this way.
The dependence on hour angle could be found explicitly
by analyzing not cataloged mean radial velocities,
but individual observations; however, this is possible
only in collaboration between all the authors of the
publications.

\begin{table}
\caption[]{Work list of secondary standards used tomake the
PCRV (the full table is presented in electronic form).}
\label{new}
\[
\begin{tabular}{rrrrrrr}
\hline
\noalign{\smallskip}
 HIP & HD & $V_r$ & $\sigma(V_r)$ & $\alpha$ & $\delta$ & $m_V$ \\
\hline
\noalign{\smallskip}
 379 & 225216 & $-28.9$ & 0.2 & 1.175 &  $67.166$ & 5.68 \\
 400 & 225261 &   $7.5$ & 0.2 & 1.235 &  $23.270$ & 7.82 \\
 428 &        &  $-0.2$ & 0.2 & 1.295 &  $45.787$ & 9.95 \\
 544 &    166 &  $-6.5$ & 0.1 & 1.653 &  $29.022$ & 6.07 \\
 616 &    283 & $-43.1$ & 0.2 & 1.886 & $-23.819$ & 8.70 \\
\hline
\end{tabular}
\]
\end{table}

The dependences on Galactic coordinates seem
to be physically unjustified. Therefore, in all of the
cases where they showed up, we sought for the dependences
on equatorial coordinates. The unique relationship
between the Galactic and equatorial coordinates
helps to choose a fit in the case where the stars
are distributed in Galactic latitude or longitude more
uniformly than they are in $\alpha$ or $\delta$.

As a result, the list of independent factors includes
the radial velocity, the $(B-V)$ color index, $\alpha$, $\delta$, the
parallax, and the $V$ magnitude. Linear and quadratic
dependences of the radial-velocity differences on the
radial velocity itself and $(B-V)$ as well as linear and
sinusoidal dependences on $\alpha$ and $\delta$ were found in the
publications under consideration.

As an example, for two publications, GCS and
de Medeiros and Mayor (1999), radial-velocity differences
of the form ``publication minus IAU list of
standards'' are plotted against $(B-V)$ , radial velocity,
and $\alpha$ in Fig. 1 for 59 and 29 standards (diamonds and
squares, respectively) before (a) and after (b) applying
the corrections from Table 3 (the lines in the Fig. 1
are shown for clarity and do not correspond to the
fits). Both publications present CORAVEL data, but
for different stars, main-sequence and giant ones. The
systematic dependences are clearly seen to be similar.

After applying the corrections, we used the 12 publications
under consideration to compile a work list of
secondary radial-velocity standards (below referred
to as the WLSS). It includes 1128 stars that meet
the following conditions: (1) a star is presented at
least in 3 of these 12 publications; (2) the accuracy of
the weighted mean velocity is higher than 1 km s$^{-1}$;
(3) the standard deviation of the results from different
publications is less than 3 km s$^{-1}$; (4) for GCS stars,
the standard deviation of the results of individual
observations is less than 5 km s$^{-1}$, while the probability
that this scatter was caused by observational
errors and not by radial-velocity variations is higher
than 0.1 (the values were taken from the GCS);
(5) a star is marked as a spectroscopic binary with
radial-velocity variations larger than 0.5 km s$^{-1}$
neither in any of the publications under consideration
nor in the Ninth Catalogue of Spectroscopic Binary
Orbits by Pourbaix et al. (2004); (6) a star does not
belong to binaries of the O, X, and G categories
from Hipparcos (the catalog's field H59); (7) a star
is not a visual pair with a component separation of
less than 20 AU (this value was calculated from the
angular separation between the components and the
Hipparcos parallax).

\begin{table}
\caption[]{Radial-velocity publications with a sufficient number of IAU and WLSS standard stars.}
\label{pubnew}
\[
\begin{tabular}{lrrrrr}
\hline
\noalign{\smallskip}
 Reference  &   N   &  n & $\overline{\Delta V_r}$ & $\sigma$ & $\sigma'$ \\
\hline
\noalign{\smallskip}
Fehrenbach et al. (1997)        & 1525 & 42 & $+4.3$ & 9.8 & 9.8   \\
Fehrenbach et al. (1992)        & 1166 & 18 & $-8.2$ & 14.4 & 11.8 \\
Duflot et al. (1992)            & 819 & 21 & $-1.7$ &  6.2 & 4.8  \\
Reid et al. (1995)              & 658 & 104 & $-2.7$ & 16.8 & 16.8  \\
Yoss and Griffin (1997)         & 631 & 19 & $+0.0$ & 0.6 & 0.6  \\
Grenier et al. (1999a)          & 547 &  6 & $-0.2$ & 1.1 & 1.1 \\
Duflot et al. (1995a)           & 535 & 10 & $-1.1$ & 3.4 & 3.4 \\
Tokovinin and Smekhov (2002)    & 424 & 20 & $+0.0$ & 0.5 & 0.5 \\
Gizis et al. (2002)             & 413 & 98 & $-0.1$ & 1.1 & 1.1 \\
Rastorguev and Glushkova (1997) & 358 &  8 & $+0.2$ & 0.6 & 0.6 \\
Ryan and Norris (1991)          & 143 & 19 & $-1.2$ & 10.1 & 10.1 \\
Abt and Willmarth (1994)        & 139 & 14 & $-0.2$ & 0.7 & 0.7  \\
Cutispoto et al. (2002)         & 104 &  5 & $-0.1$ & 0.2 & 0.2 \\
King et al. (2003)              &  86 &  7 & $+0.5$ & 1.0 & 1.0 \\
Tokovinin (1990)                &  85 & 32 & $-0.1$ & 0.3 & 0.3 \\
Marrese et al. (2003)           &  83 & 21 & $-0.4$ & 1.0 & 1.0 \\
Griffin and Suchkov (2003)      &  70 &  5 & $+0.3$ & 0.6 & 0.6 \\
Clementini et al. (1999)        &  62 & 29 & $+0.0$ & 0.8 & 0.8 \\
Delfosse et al. (1998)          &  60 & 45 & $+0.0$ & 0.4 & 0.4 \\
Garcia-Sanchez et al. (1999)    &  58 & 22 & $-0.1$ & 0.3 & 0.3 \\
Morrell and Abt (1992)          &  43 &  6 & $-0.2$ & 1.2 & 1.2 \\
Soderblom and Mayor (1993)      &  37 &  8 & $+0.1$ & 0.1 & 0.1 \\
Gaidos et al. (2000)            &  34 & 17 & $+0.1$ & 0.3 & 0.3 \\
Metzger et al. (1992)           &  32 &  8 & $+0.0$ & 0.7 & 0.7 \\
Takeda et al. (1998)            &  32 & 15 & $+0.3$ & 0.6 & 0.6 \\
Qiu et al. (2002)               &  24 & 10 & $+0.3$ & 0.5 & 0.5 \\
Olszewski et al. (1995)         &  20 &  8 & $+0.2$ & 0.2 & 0.2 \\
Orosz et al. (1997)             &  13 & 12 & $-0.7$ & 0.7 & 0.4 \\
Mazeh et al. (2002)             &  13 &  6 & $-1.4$ & 1.6 & 1.6 \\
Gonzalez et al. (2001)          &  12 & 12 & $+0.1$ & 0.4 & 0.2 \\
Metzger et al. (1991)           &  12 & 10 & $+0.4$ & 0.4 & 0.4 \\
Dubath et al. (1997)            &  10 & 10 & $+0.5$ & 1.2 & 1.2 \\
Metzger et al. (1998)           &   9 &  9 & $+0.3$ & 1.5 & 0.8 \\
\hline
\end{tabular}
\]
\end{table}

The WLSS is presented in Table 4: it gives Hipparcos
and HD numbers, weighted mean radial velocities
with their accuracies $\sigma(V_r)$ in the PCRV (in
km s$^{-1}$), approximate $\alpha$ and $\delta$ in degrees with fractions
(J2000), and $V$ magnitudes.

The distributions of the 155 IAU standards and
1128 WLSS stars under consideration over the celestial
sphere are indicated in Fig. 2 by diamonds and
crosses, respectively. The stars are distributed uniformly
in $\alpha$ and predominate in the Northern Hemisphere.
The distributions of the same stars in $(B-V)$
color index and absolute magnitudeMV are indicated
in Fig. 3 by the same symbols. We see that, in contrast
to the IAU standards, the WLSS includes red
dwarfs, subgiants, and main-sequence stars in the
interval A5-F5. However, this interval, with a large
number of peculiar and rapidly rotating stars, remains
a problem in radial-velocity measurements. In addition,
there are no supergiants, white dwarfs, and
O-type stars in the WLSS. Of considerable interest
are the distributions of the same stars in distance, as
derived from the Hipparcos parallax, and apparent $V$
magnitude, which are indicated in Fig. 4 by the same
symbols. Here, we clearly see the division of stars into
three groups: (1) nearby red dwarfs from the WLSS,
(2) F5V-G8V stars from both lists at intermediate
distances, and (3) distant red giants together with
early-type stars with a predominance of IAU standards.

\begin{table}
\caption[]{Systematic radial-velocity corrections for publications with a sufficient number of IAU and WLSS standard
stars.}
\label{cornew}
\[
\begin{tabular}{lc}
\hline
\noalign{\smallskip}
 Reference  & Correction \\
\hline
\noalign{\smallskip}
Fehrenbach et al. (1997)        & $              -4.3                     $ \\
Fehrenbach et al. (1992)        & $ -38.88(B-V)-0.369\delta+36.64         $ \\
Duflot et al. (1992)           & $ -4.05\cos(\alpha)-0.0114\delta^{2}+1.16\delta-23.2 $ \\
Reid et al. (1995)           & $                     +2.7              $ \\
Yoss and Griffin (1997)         & $                      0                $ \\
Grenier et al. (1999a)         & $                     +0.2              $ \\
Duflot et al. (1995a)          & $                     +1.1              $ \\
Tokovinin and Smekhov (2002)     & $                      0                $ \\
Gizis et al. (2002)          & $                     -0.2              $ \\
Rastorguev and Glushkova (1997) & $                      0                $ \\
Ryan and Norris (1991)         & $                      +1.2             $ \\
Abt and Willmarth (1994)         & $                     +0.2              $ \\
Cutispoto et al. (2002)       & $                     +0.1              $ \\
King et al. (2003)           & $                     -0.5              $ \\
Tokovinin (1990)            & $                     +0.1              $ \\
Marrese et al. (2003)         & $                     +0.4              $ \\
Griffin and Suchkov (2003)      & $                     -0.3              $ \\
Clementini et al. (1999)     & $                      0                $ \\
Delfosse et al. (1998)         & $                      0                $ \\
Garcia-Sanchez et al. (1999)  & $ -0.37\cos(2(\alpha-30))               $ \\
Morrell and Abt (1992)         & $                     +0.2              $ \\
Soderblom and Mayor (1993)     & $                    -0.1               $ \\
Gaidos et al. (2000)         & $                     -0.1              $ \\
Metzger et al. (1992)         & $                      0                $ \\
Takeda et al. (1998)         & $                     -0.3              $ \\
Qiu et al. (2002)            & $                      -0.3             $ \\
Olszewski et al. (1995)      & $ +0.005V_{r}-0.17                      $ \\
Orosz et al. (1997)           & $ -1.99(B-V)+0.0145V_{r}-3.0            $ \\
Mazeh et al. (2002)          & $                    +1.4               $ \\
Gonzalez et al. (2001)       & $ -0.56\sin(\alpha)-0.002\alpha+0.29    $ \\
Metzger et al. (1991)         & $                      -0.4             $ \\
Dubath et al. (1997)           & $                      -0.5             $ \\
Metzger et al. (1998)        & $ -2.12\cos(\alpha)-1.2                 $ \\
\hline
\end{tabular}
\]
\end{table}

We used the combined list of 1283 standard stars
(IAU+WLSS) to estimate and correct the radial velocity
zero points and systematic errors in 33 publications
with the results of observations for these
stars. For these publications, Tables 5 and 6 give
characteristics similar to those presented in Tables 2
and 3, but the combined list of standards is used
instead the IAU list of standards. We used 8258 mean
radial velocities from these publications, which accounted
for $\sim13\%$ of the PCRV material. The analysis
of this material, the list of factors, and the procedure
for calculating the weights and weighted mean
radial velocities are similar to those described above
for the 12 major publications. No significant dependences
on any factors were found for most of these
publications; only the zero point was corrected.

In making the PCRV, we considered a total of
more than 1000 publications with radial-velocity
measurements. The full list is accessible at the OSACA
page (http://www.geocities.com/orionspiral/).
However, most of them pertain to single observations
of individual stars that are not only nonstandard ones,
but generally belong to spectroscopic binaries or
peculiar stars. In such cases, the systematic errors
of the derived radial velocities are difficult to estimate.
Nevertheless, we selected 154 publications among
those that did not include IAU and WLSS standards
for which rough estimates of the systematic errors
can be given. These publications were produced with
the same instruments as those considered above
or the publications themselves provide convincing
evidence that the systematic errors of the derived
radial velocities do not exceed 1 km s$^{-1}$. The weights
were assigned on this basis. The radial velocities from
these publications were used in the PCRV without
corrections. The largest of these 154 publications
are presented in Table 7: it gives a reference to
the publication, the number N of stars from the
publication used in the PCRV, and an estimate of
the accuracy in assigning the weight $\sigma$. We analyzed
4181 mean radial velocities from these publications,
which accounts for $\sim7\%$ of the PCRV material.

\begin{table}
\caption[]{Some publications with radial velocities used in the PCRV without corrections.}
\label{pubrest}
\[
\begin{tabular}{lrr}
\hline
\noalign{\smallskip}
  Reference     &  N  & $\sigma$ \\
\hline
\noalign{\smallskip}
Nordstr\"om et al. (1997) & 574 & 0.3 \\ 
Griffin (2006)        & 207 & 0.4 \\ 
deMedeiros et al. (2002)  & 178 & 0.2 \\ 
Carrier et al. (2002)    & 108 & 2.7 \\ 
Behr (2003)            &  95 & 1.4 \\ 
deMedeiros et al. (2004)  &  94 & 1.2 \\ 
Carney et al. (2003)    &  84 & 0.3 \\ 
Gorynya et al. (1996)   &  72 & 0.3 \\ 
\hline
\end{tabular}
\]
\end{table}

We analyzed a total of 60 207 mean radial velocities
for 40 825 stars. However, the PCRV contains
only those of the numerous spectroscopic binaries
for which the radial center-of-mass velocity
(systemic velocity) can be calculated. As a rule, these
are orbital pairs from the Ninth Catalogue of Spectroscopic
Binary Orbits by Pourbaix et al. (2004).
In addition, we found several hundred cases where
the same component was designated differently in
different publications or different components were
designated identically. As a result, more than 5000
``problem'' stars were put off until the publication
(presumably in 2006) of the Washington Multiplicity
Catalog (WMC), in which a unique designation will
probably be given for the components of nonsingle
stars.

There are two compilation catalogs among the
publications used (see Table 2): WEB and BBF. As
we see from the table, they are fairly accurate, at least
with regard to the radial velocities of IAU standard
stars. The distribution of radial-velocity differences of
the form ``publication minus IAU list of standards''
is nearly Gaussian. The accuracies declared by the
authors of these publications may be considered plausible.
These rather than the much more ``optimistic''
values of $\sigma'(\Delta V_r)$ from Table 2 were used to calculate
the weights. The unexpectedly high accuracy of these
publications probably stems from the fact that they
are actually based on the results of a few scientific
groups and instruments: for example, the 48 largest
catalogs of the 459 initial catalogs used in BBF include
75\% of the data and they were compiled by
no more than 20 scientific groups. In the future, the
material from these publications is planned to be used
on the level of initial observational catalogs.

\section*{THE PULKOVO COMPILATION OF RADIAL VELOCITIES}

The Pulkovo Compilation of Radial Velocities
(PCRV) is presented in Table 8: it gives Hipparcos
and HD numbers, weighted mean radial velocities $V_r$
in km s$^{-1}$, their accuracies $\sigma_{int}$ in km s$^{-1}$ (calculated
using the standard formula $1/(\sum p)^{-1/2}$, where $\sum p$ is
the sum of the weights), the number $n$ of publications
used, the radial-velocity accuracy $\sigma_{ext}$ calculated as
the standard deviation of the differences ``publication
minus weighted mean velocity'' for the case of two or
more publications in km s$^{-1}$, approximate $\alpha$ and $\delta$
(J2000) in degrees with fractions, and Hipparcos $V$ magnitudes.

The distribution of 35 495 PCRV stars over the
celestial sphere is shown in Fig. 5: a predominance of
stars in the Northern Hemisphere and a concentration
to the Galactic equator and the Galactic North
Pole are clearly seen. The distribution in $(B-V)$ color
index and absolute magnitudeMV is shown in Fig. 6.
This distribution is are the same as that for all Hipparcos
stars: all classes are represented. The distribution
in distance, as derived from the Hipparcos parallax,
and apparent magnitude $V$ is shown in Fig. 7. About
30 000 (85\%) stars lie at distances from 30 to 500 pc.

The median accuracy of the weighted mean radial
velocity in the PCRV is 0.7 km s$^{-1}$. The radial
velocities of 21 015 (59\%) and 29 804 (84\%) stars
have accuracies higher than 1 and 3 km s$^{-1}$, respectively.
Unfortunately, the stars whose radial velocities
are given in only one publication predominate,
i.e., 24 437 (69\%) stars. The radial velocities of 8319
(23\%) and 2739 stars are given in 2 and from 3
to 11 publications, respectively. The median of $\sigma_{ext}$,
the standard deviation of the differences ``publication
minus weighted mean velocity'' for the case of two or
more publications, is 1.5 km s$^{-1}$. Since hundreds of
spectroscopic binaries may be hidden in the PCRV,
we conclude that the systematic errors of the publications
used have been successfully taken into account.

\begin{table}
\caption[]{Pulkovo Compilation of Radial Velocities (the full table is presented in electronic form).}
\label{pcrv}
\[
\begin{tabular}{rrrrrrrrr}
\hline
\noalign{\smallskip}
 HIP & HD & $V_{r}$ & $\sigma_{int}$ & n & $\sigma_{ext}$ & $\alpha$ & $\delta$ & $m_{V}$  \\
\hline
\noalign{\smallskip}
  3 & 224699 &   $0.0$ & 4.2 & 1 &     & 0.005 & 38.859 & 6.61 \\
  7 &        &   $8.3$ & 1.5 & 1 &     & 0.022 & 20.036 & 9.64 \\
  8 & 224709 & $-31.0$ & 4.6 & 1 &     & 0.027 & 25.886 & 9.05 \\
 11 & 224720 & $-25.8$ & 2.4 & 1 &     & 0.037 & 46.940 & 7.34 \\
 14 & 224726 &  $21.7$ & 0.9 & 1 &     & 0.048 & $-0.360$ & 7.25 \\
 19 & 224721 &   $6.3$ & 0.7 & 2 & 9.7 & 0.053 & 38.304 & 6.53 \\
\hline
\end{tabular}
\]
\end{table}

\section*{CONCLUSIONS}

The Pulkovo Compilation of Radial Velocities
(PCRV) for 35 495 Hipparcos stars has been made
by analyzing more than 60 000 mean radial velocities
from 203 publications. Owing to the observations of
155 IAU standard stars, 12 publications containing
$\sim80\%$ of all the radial velocities used were freed with
confidence from systematic errors and errors in the
radial-velocity zero points that depend on the radial
velocity itself, $(B-V)$ color index, and equatorial coordinates.
Based on the data from these publications,
we compiled a work list of 1128 secondary standards.
This list is an important result of our study, since it
is considerably better than the IAU list, represents
the whole variety of stellar classes, includes many
faint stars, and covers more uniformly the celestial
sphere. This allows it to be used to analyze and
evaluate new radial-velocity observations. Our study
has shown that the increasing flow of current radial velocity
measurements cannot be analyzed without
producing such a list of secondary standards. The
two lists of standards were used together to free
33 more publications containing $\sim13\%$ of all the
radial velocities used from systematic errors and
errors in the radial-velocity zero points. The objective
estimates of the accuracy of the publications obtained
from comparison with the lists of standards allow
the median estimate of the PCRV accuracy to be
considered plausible: 0.7 km s$^{-1}$. All the main classes
stars over the entire celestial sphere mostly within
500 pc of the Sun are presented in the PCRV.

The derived radial velocities, together with the
Hipparcos three-dimensional coordinates, Tycho-2
propermotions, photometry, and data on the duplicity,
chemical compositions, and ages of stars in the form
of OSACA, are already used in kinematic studies. For
example, Gontcharov and Vityazev (2005) showed
that although many Galactic structures within 400 pc
of the Sun (the Local Bubble, the Great Tunnel,
Gould's Belt) consist of stars of different ages up
to several hundred million years, they were formed
recently, no earlier than 20 Myr ago in processes
related to two stellar streams, Orion and Sirius.
Based on the Milne-Ogorodnikov model, Bobylev
et al. (2006) have demonstrated significant differences
in the kinematics of the nearest single and multiple
main-sequence stars and accurately determined the
Galactic rotation parameters from the motions of
stars farther than 200 pc.

\section*{ACKNOWLEDGMENTS}

I wish to thank R. Griffin and A.S. Rastorguev
for advice, consultations, and unpublished materials.
I am grateful to the anonymous referee for helpful remarks.
In this study, I have intensively used the SIMBAD
Astronomical Database and other resources
from the Astronomical Data Center in Strasbourg
(France) (http://cdsweb.u-strasbg.fr/). This study
was supported by the Russian Foundation for Basic
Research, project no. 05-02-17047.

\newpage

\begin{figure}
\includegraphics{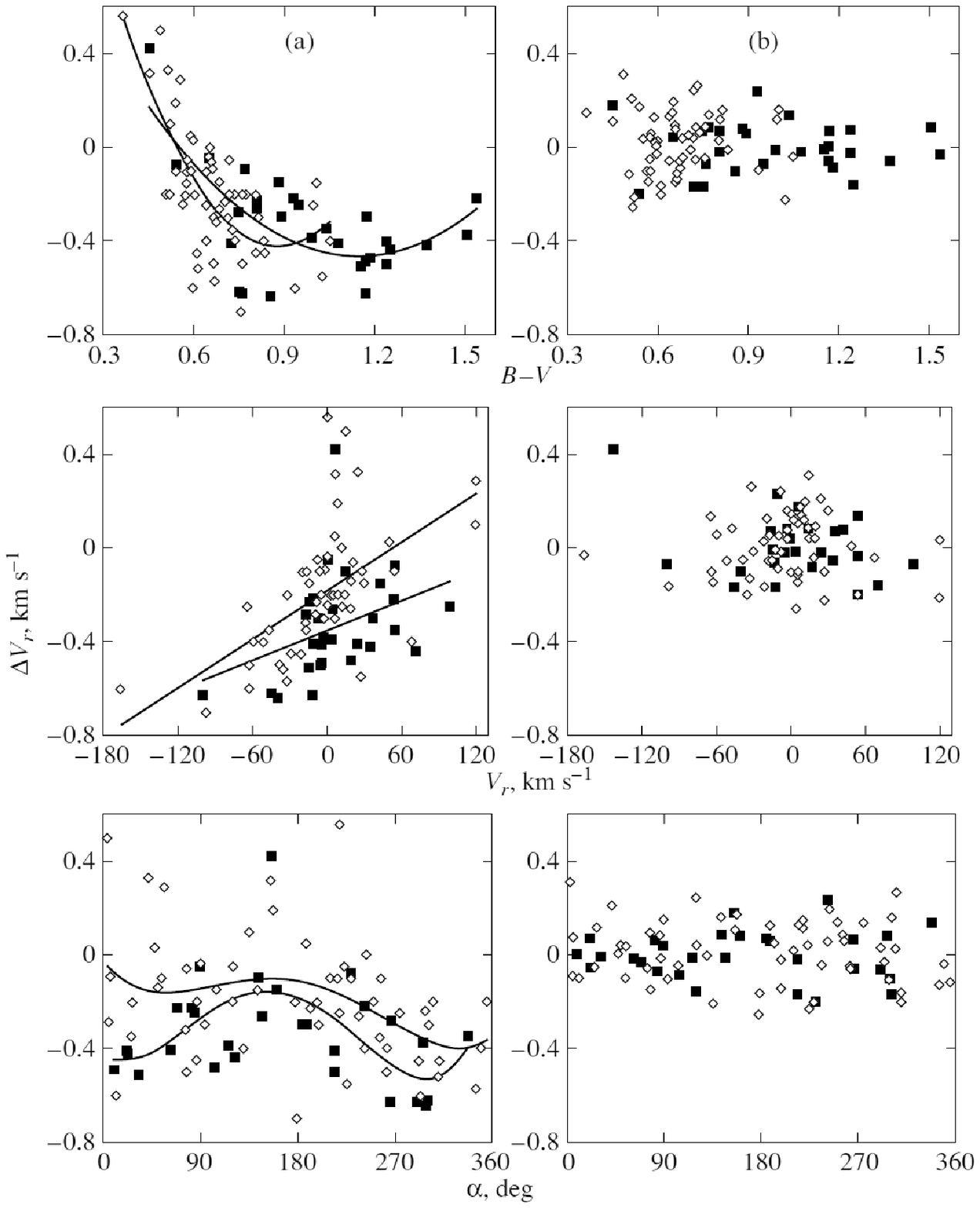}
\caption{Radial-velocity differences of the form ``publication minus IAU list of standards'', vs. $(B-V)$ color index,
radial velocity, and $\alpha$ before and after applying the corrections from Table 3 for the GCS and the catalog by
de Medeiros and Mayor (1999), diamonds and squares, respectively, for 59 and 29 IAU standard stars.}
\label{coravel}
\end{figure}

\begin{figure}
\includegraphics{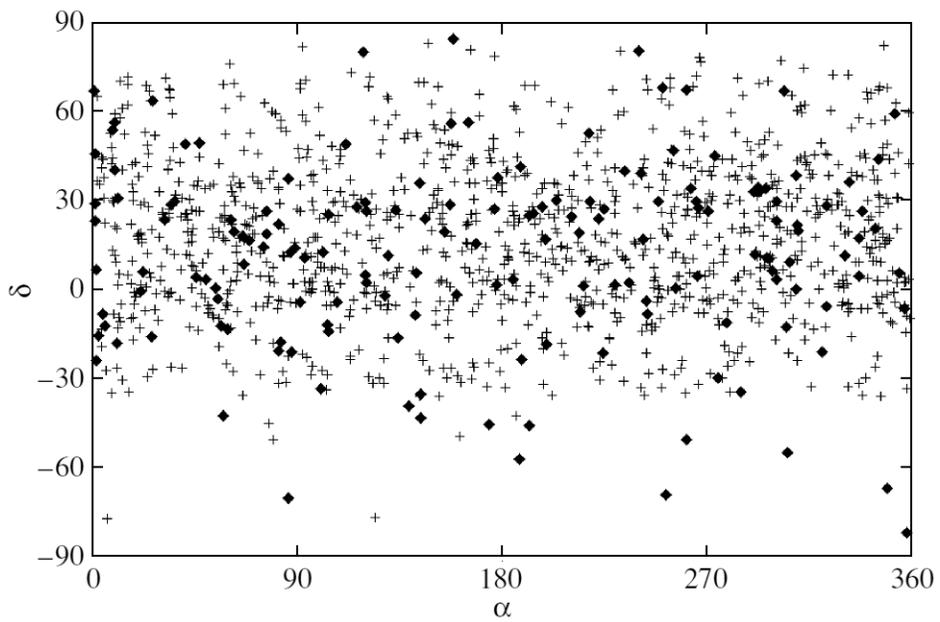}
\caption{Distribution of 155 IAU standards (diamonds) and 1128 WLSS stars (crosses) over the celestial sphere in
equatorial coordinates.}
\label{srd}
\end{figure}

\begin{figure}
\includegraphics{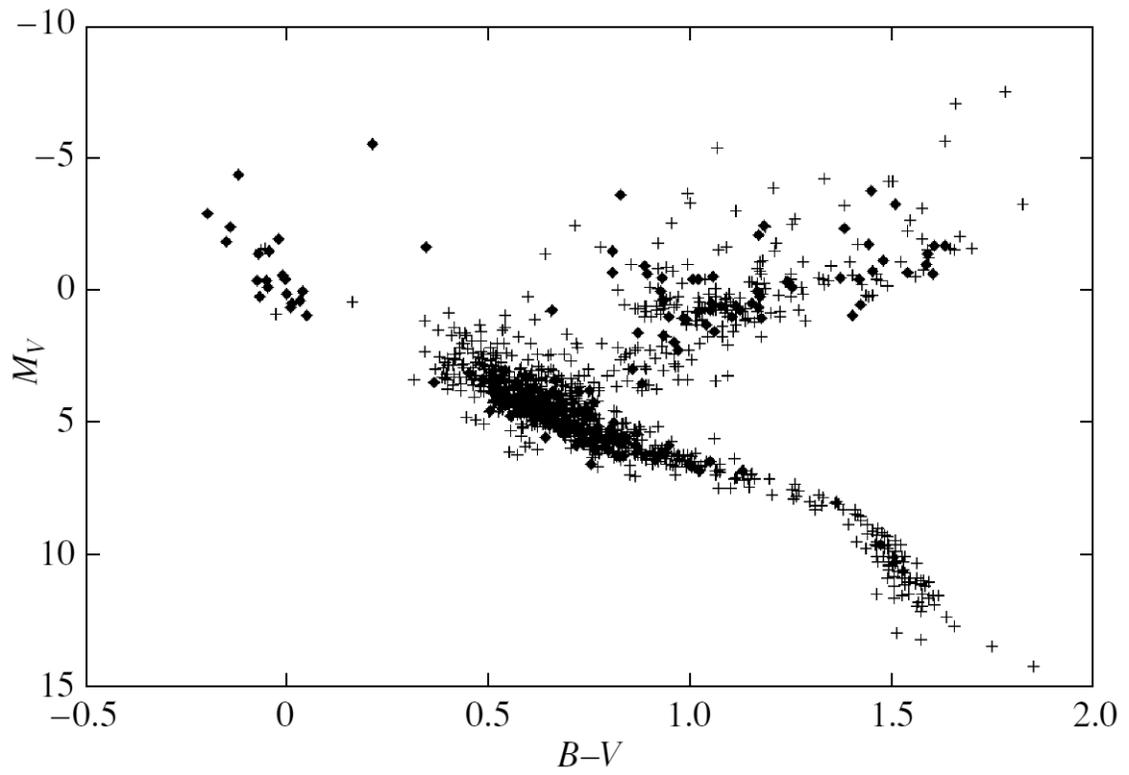}
\caption{Distribution of 155 IAU standards (diamonds) and 1128 WLSS stars (crosses) in $(B-V)$ color index and absolute
magnitude.}
\label{shr}
\end{figure}

\begin{figure}
\includegraphics{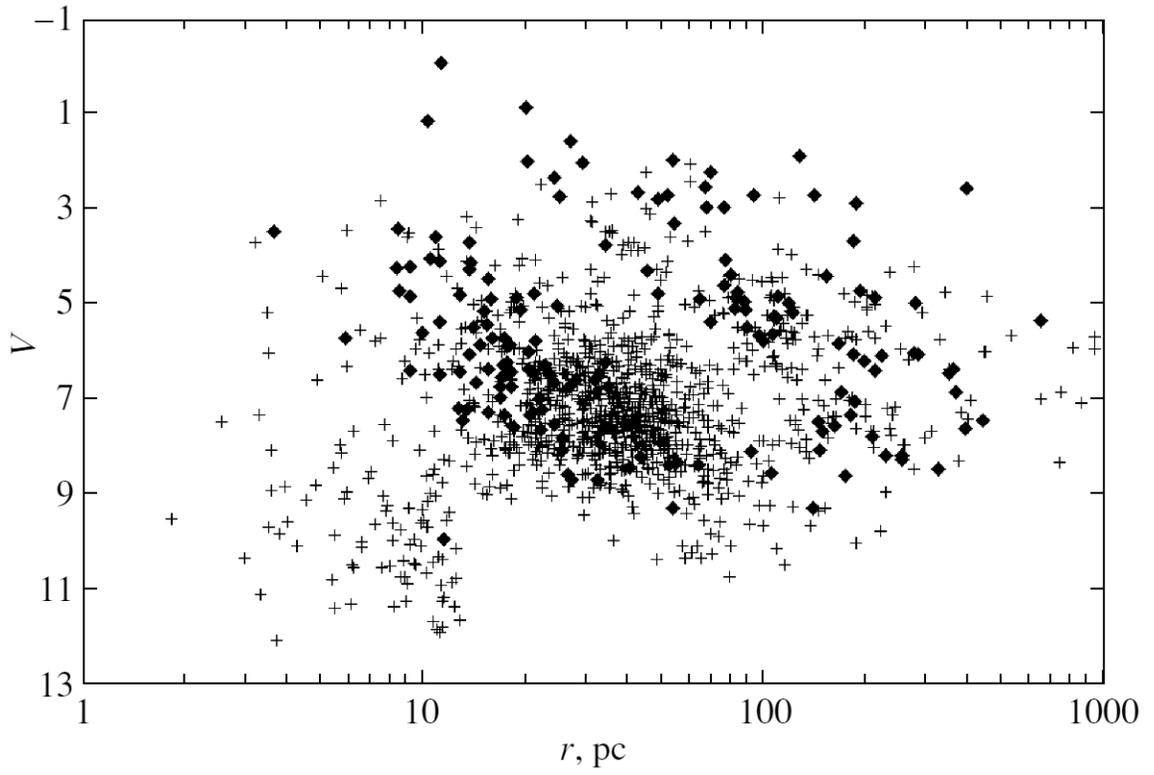}
\caption{Distribution of 155 IAU standards (diamonds) and 1128 WLSS stars (crosses) in distance and $V$ magnitude.}
\label{srm}
\end{figure}

\begin{figure}
\includegraphics{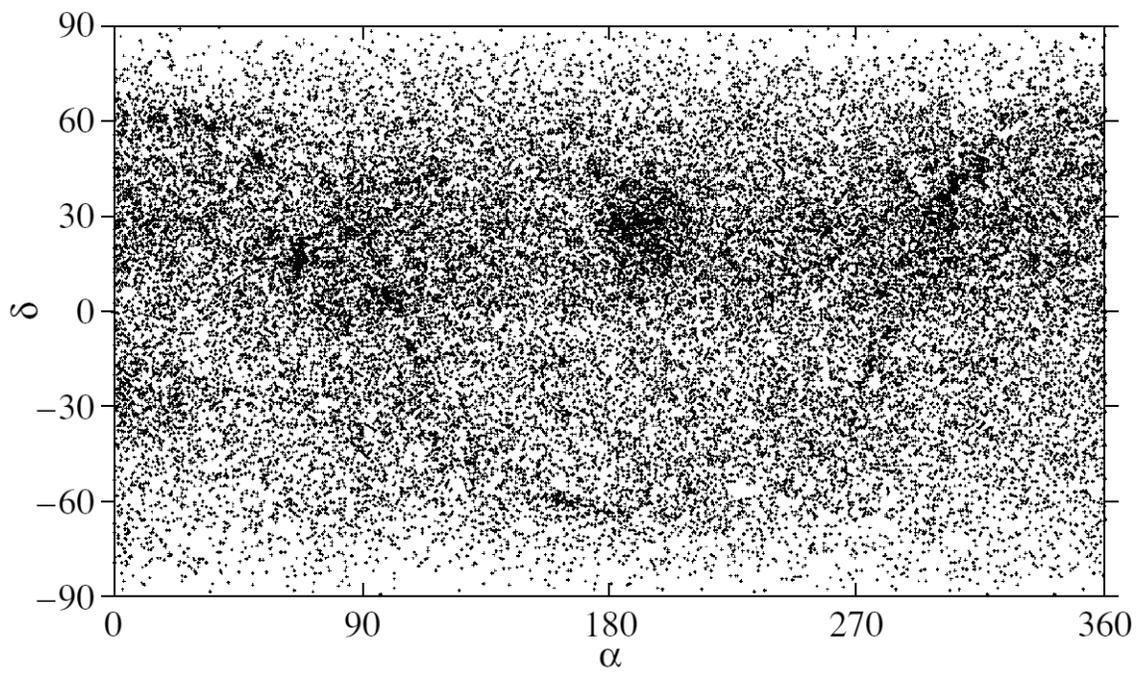}
\caption{Distribution of 35 495 PCRV stars over the celestial sphere in equatorial coordinates.}
\label{rd}
\end{figure}

\begin{figure}
\includegraphics{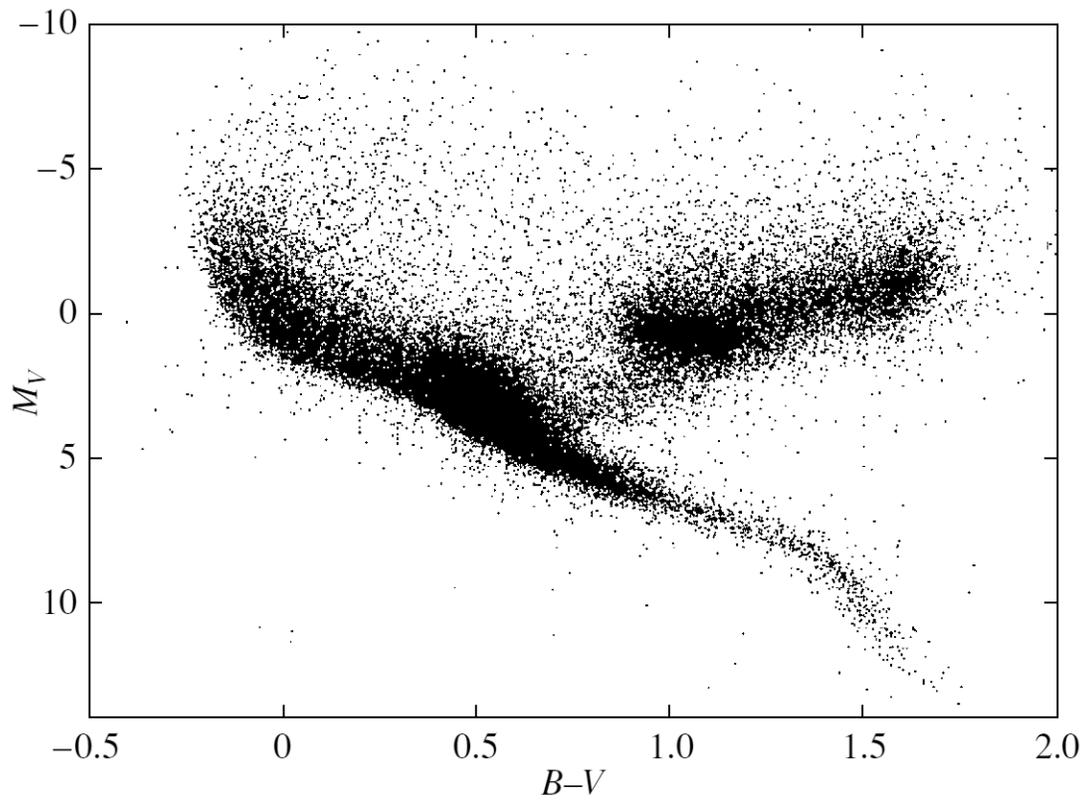}
\caption{Distribution of 35 495 PCRV stars in $(B-V)$ color index and absolute magnitude.}
\label{hr}
\end{figure}

\begin{figure}
\includegraphics{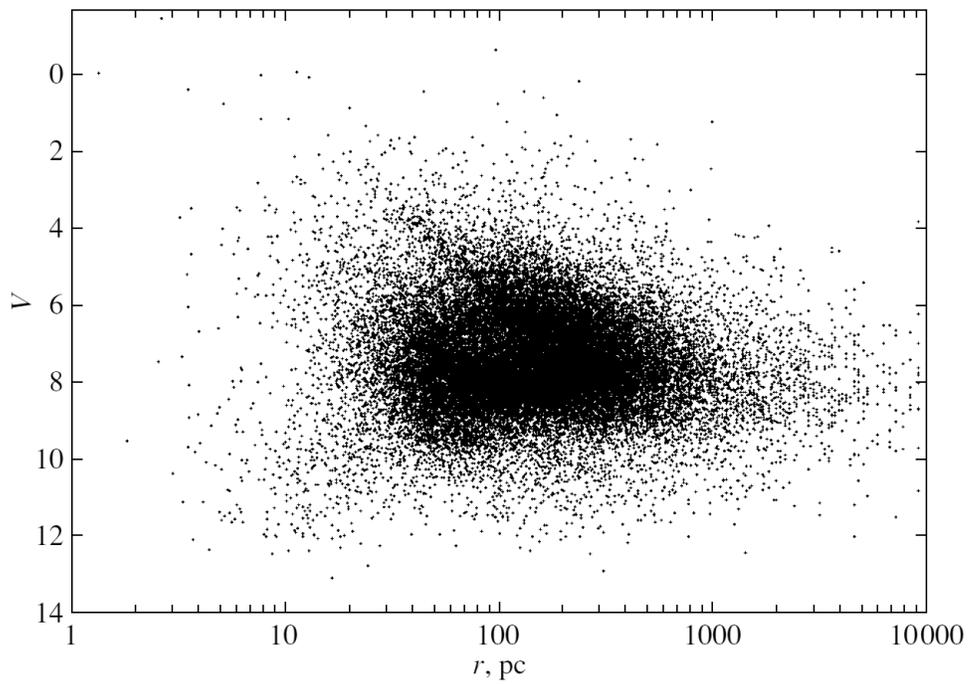}
\caption{Distribution of 35 495 PCRV stars in distance and $V$ magnitude.}
\label{rm}
\end{figure}

\end{document}